\documentstyle[twoside,fleqn,espcrc2,epsf]{article}

\newcommand{\beq}{\begin{equation}}
\newcommand{\eeq}{\end{equation}}

\newcommand{\AmS}{{\protect\the\textfont2
  A\kern-.1667em\lower.5ex\hbox{M}\kern-.125emS}}

\title{Lattice Quantum Gravity :\\
       Review and Recent Developments}

\author{S. Catterall\address{Department of Physics, 
        Syracuse University,\\ 
        Syracuse NY 13244}}
       
\begin{document}

\begin{abstract}
We review the status of different approaches to lattice quantum gravity
indicating the successes and problems of each. Recent developments 
within the dynamical triangulation formulation are then described.
\end{abstract}

\maketitle

\section{Introduction}

The problem of uniting the two pillars of twentieth century
physics, namely general relativity and quantum mechanics, is 
arguably one of the most important and challenging goals of
theoretical physics. Much effort has been expended on the task
of formulating such a {\it quantum gravity} theory extending
over a period of several decades. In this review we want
to concentrate on the status of a relatively new approach to
the problem based on replacing continuum spacetimes by simplicial
approximations.

We will be working within a path integral approach in which 
the partition function for gravity is gotten by performing
a functional integral over an appropriate set of spacetime
geometries

\beq
Z=\int D\left[g\right] e^{-S\left(g\right)}.
\eeq

The evaluation of this partition function presents several, well-known
problems. Firstly, if we choose to work with positive definite metrics
(Euclidean space), the Einstein-Hilbert 
action $S\left(g\right)$ is unbounded from
below due to the conformal mode. Second, there exists no unique,
physically well-motivated choice for the functional measure
$D\left[g\right]$. Clearly at minimum the measure should
ensure that we sum only over physically inequivalent metrics - those
not related by reparametrizations. 

The third problem relates to
the use of perturbation theory - the Newton constant has mass
dimension minus two which ensures that the model is
not perturbatively renormalizable. Indeed, taken at
face value this implies that fluctuations of
the geometry become ever more severe at smaller
and smaller distance scales and it is not even clear
that a continuum spacetime is an appropriate way of
describing it. Finally, the sum over
geometries should presumably include manifolds with varying
topology. Unfortunately, there is no scheme for classifying
four-manifolds by their topology unlike the situation in two
dimensions. This means it is not even clear how to
formulate such a sum over topologies. Hence we will ignore
this problem throughout the rest of this talk.

Various methods have been proposed \cite{qgrav} to evade some or all of
these problems - canonical methods have been devised to
try to avoid the use of perturbation theory, supersymmetric
extensions of the Einstein theory have been considered in the hope
of cancelling loop divergences. In string theory (the
most radical solution so far) the Einstein-Hilbert
theory emerges as a low energy effective field theory. However,
none of these approaches has been entirely successful and so
another approach has been proposed based on a lattice
approximation which we will now discuss.

\section{Simplicial Quantum Gravity} 

The original idea of using simplicial lattices to
approximate smooth manifolds is due to Regge \cite{regge}.
The idea is that any manifold can be viewed locally
as being composed of pieces of flat space appropriately
glued together. In $d$ dimensions it is convenient to use as elementary
building blocks $d$-simplices which are assembled together by
identifying pairs of $(d-1)$ dimensional faces. A 
(sub)simplex of size $d$ is composed of $(d+1)$ labels
corresponding to its vertices. In two
dimensions such objects are triangles, in three dimensions
tetrahedra and in four hypertetrahedra. 
A simplicial complex $K$ or triangulation is a set of such
simplices together with a specific gluing. It is usual
to impose a `manifold' condition on this gluing; the neighborhood
of any point should be homeomorphic to a $d$-dimensional ball.
This effectively eliminates triangulations possessing 
degenerate subsimplices.

Two sorts of degrees of freedom are now apparent; the 
edge lengths of the basic simplices $\{l_i\}$ and the triangulation
$T$ itself.
Using these variables Regge was able to write down
discrete equivalents of the Einstein action, the volume
element etc.
While Regge's original suggestion was confined to the
solution of {\it classical} problems in general
relativity it is natural to try to generalize it
to the quantum domain by constructing a lattice partition
function

\beq
Z_L=\sum_{K\epsilon\left(T,\left\{l_i\right\}\right)}
\rho\left(T,l\right)e^{-S_L\left(T,l\right)}.
\label{zl}
\eeq

In practice this partition function is still difficult to
handle and two further approaches have been taken;
Regge Quantum Gravity (RQG) and Dynamical Triangulation (DT).
In the former, the triangulation is held fixed and one simply
integrates over link lengths $l_i$. The second approach treats
all the link lengths as equal to some reparametrization invariant
cutoff $a$ and a sum is performed over all triangulations $T$.

\section{Successes and Difficulties}

\subsection{RQG}

First consider the RQG formulation. Its primary advantage is that
it is possible to show that its lattice action reduces to
the continuum action in some appropriate limit in which
the mean simplex edge length is reduced to zero as the 
number of simplices  is taken to infinity. It similarly
possesses a weak field expansion in which contact can be
made with continuum perturbation theory - rather like
the naive weak coupling limit of SU(3) lattice
gauge theory. While this is encouraging it is of course
insufficient for proving that we are really studying a
quantum theory of gravity outside of perturbation theory.  We will
see evidence for this in the numerical results that have been
obtained in two dimensions.

One of the most exciting features of this approach has
been the observation of a candidate continuous phase
transition in the four-dimensional theory \cite{hamber,riedler}.
Current efforts focus on analyzing the scaling
behavior of this model in the vicinity of
the phase transition and extracting estimates for critical 
exponents. 

The problems with this approach are four-fold. Firstly there are
no exact solutions of the full quantum problem which can
tested against the predictions of continuum calculations. In two
dimensions several models incorporating critical matter fields
coupled to quantum gravity have been solved using CFT 
techniques. Unfortunately the RQG model cannot be solved 
exactly here to make detailed comparisons. In contrast we will
see that this has been done for many of the DT models with
results that show complete agreement with the continuum 
calculations.

Secondly, there remains a large ambiguity in the choice of
measure for the link length integration. Commonly this
measure is written as

\beq
D\left[g\right]\to\prod_i\int {dl_i^2\over l_i^{2\sigma}}.
\eeq

The question remains as to whether this is a suitably general
choice of measure - specifically will such a {\it local}
measure be sufficient and how should the power $\sigma$ be
chosen?

These problems are highlighted in two dimensions for the critical
Ising model coupled to gravity. A sequence of simulations has
revealed that the RQG approach simply fails to yield the
correct {\it gravitationally dressed} Ising exponents 
\cite{gross,holm}.

Finally, it has been pointed out that a naive integration
over all link lengths grossly overcounts physical degrees of
freedom by including modes corresponding to general
coordinate transformations in the continuum \cite{bock}. 
The authors argue that the correct continuum limit can only
be obtained by utilizing an appropriate
gauge fixing procedure to eliminate the contributions 
of these gauge degrees of freedom.

Recently, an analytic calculation within the RQG framework has
been completed in which the Fadeev-Popov determinant is
computed for two-dimensional pure gravity \cite{menotti}.
The authors find that the Fadeev-Popov determinant when carefully
regulated produces a non-local effective action which
{\it precisely} matches the continuum conformal anomaly. It is
the latter which is responsible for the dressing of
the exponents. Any calculation which leaves this
term out by not gauge fixing has no hope of generating the
correct physics in two dimensions. It is not clear, however
that the situation is necessarily so serious in higher
dimensions.

\subsection{DT}

The inspiration for this approach really arose out of
attempts to write down discrete regularizations for string 
theories; the latter being equivalent to two-dimensional
quantum gravity coupled to a variety of matter fields
\cite{dav,amb,kaz}. Much of this work had been anticipated
earlier by Weingarten in models of random
surfaces embedded in hypercubic lattices \cite{wein}.

One of the primary advantages of this approach is that in two
dimensions a variety of exact solution exist which may be compared
with the solutions of continuum Liouville theory. A remarkable
result has emerged from these studies; the DT correlation
functions computed on arbitrary genus graphs agree with
the corresponding quantities computed with continuum CFT
techniques.

Furthermore, the critical behavior of these models has been
studied numerically over the past few years and again very good
agreement has been seen with the theoretical predictions
\cite{generic_num}. The overall
conclusion from this work has been that, at least for dimension two,
the DT method affords a good prescription for regulating quantum
gravity.  

The potential problems for this approach in part derive from
its nonperturbative formulation as a sum over all
triangulations with equal weight - there is no weak coupling
limit in which contact can be made with continuum
perturbation theory. Indeed, the attractive feature of
this formulation - that it is purely geometric making no
reference to coordinates and metric tensors also poses
a problem; how do such classical quantities emerge from the
model at large distance. Specifically we can ask the
question: how do we recover the continuum metric tensor
from the ensemble of triangulations? These are issues whose
resolution will,
we believe, prove crucial in determining the physical
content of the higher dimensional models. The DT approach,
while intuitively appealing, must remain essentially an
ansatz for these latter models.

\section{Summary of Background}

There exist two closely related approaches to the problem
of formulating a lattice theory of quantum gravity - one
based closely on Regge's original proposal - RQG, and another
called dynamical triangulation (DT) in which the dynamical
variables correspond to abstract triangulations. 

Both have seen some success. This is particularly true for
the DT models in which a variety of exact results have
effectively proven them equivalent to the continuum 
approaches wherever the latter can be solved. In both cases
efficient numerical algorithms exist which can be used
to probe the nonperturbative content of the theories. 
Both approaches show candidate phase transitions in four
dimensions and it is an open question whether these lie
in the same universality class.

From the rest of this talk, because of time constraints, I shall
consider only the DT approach and refer the interested reader
to the parallel sessions in this meeting for recent
developments within the RQG framework \cite{beirl,berg,krishnan,nishimura}.

\section{The DT Model}

In four dimensions the DT limit of the partition function eqn.\ref{zl}
can be written
\beq
Z=\sum_{T\left(S^4\right)}e^{\kappa_0 N_0-\kappa_4 N_4},
\eeq
where the coupling $\kappa_0$ is conjugate to the total number of
vertices $N_0$ and corresponds to a bare (inverse) Newton constant. 
Likewise $\kappa_4$ is a lattice cosmological constant coupled to
the total volume or number of four-simplices $N_4$. We sum over
a class of triangulations of the four-sphere which satisfy
a certain `manifold-restriction'. namely, the boundary of the
region in $T$ enclosing any vertex is homeomorphic (`looks-like')
$S^3$. To be concrete imagine a two-dimensional triangulation and
focus on one point within that triangulation. If that point were to
be removed (together with its links to neighbors) a polygonal hole
would be created whose boundary would be a bona fide triangulation
of $S^1$ - a piecewise linear circle. 

A set of `moves' or local retriangulations of such $d$-dimensional
triangulated manifolds has long been known to exist \cite{alex} and to
satisfy an ergodic property: any triangulation can be
transformed into any other by an appropriate sequence of such
moves \cite{var_gross}.
In $d$ dimensions there are $(d+1)$ types of move which may
labelled by the order $i$ of a subsimplex central to the move. If
this randomly chosen subsimplex is associated with $(d+1-i)$ 
simplices it is possible to try to replace it with a `dual'
$(d-i)$-subsimplex provided that the manifold condition is
not violated.
By utilizing these moves inside a metropolis algorithm it is
possible to sample the dominant triangulations in the
partition function see eg \cite{simon}. This allows a wide
range of nonperturbative investigations to be carried out
which are currently inaccessible to analytic methods. We will
discuss the results of some recent numerical studies which
focus on a variety of topics; the nature of the quantum geometry, the
development of MCRG methods for DT systems and detailed studies of
the phase transition and correlation functions in four dimensions.
Finally we will discuss some of the open issues and future
research directions. 

\section{Recent Progress in Two Dimensions}

\subsection{Quantum Geometry}

There are many quantities of physical interest which are not
currently accessible to analytic study even in `exactly
solved' models such as pure $2d$ gravity. One of the most 
interesting of these concerns the nature of the quantum
geometry, for example, does it exhibit simple fractal properties,
what are the typical geodesic paths etc.

These questions may be at least partially answered by looking
at the distribution of geodesic paths as a function of lattice
volume $n\left(r,V\right)$
\beq 
n\left(r,V\right)={1\over V}\left\langle\sum_{ij}\delta\left(d_{ij}-r\right)
\right\rangle
\eeq

The function $n\left(r,V\right)$ counts the number of points which
may be reached from some
origin in $r$ steps along links of a random triangulation with volume $V$.
If $n\left(r,V\right)$ has the following simple behavior
\beq
n\left(r,V\right)\sim r^{d_H-1},\qquad V\to\infty,
\label{dist}
\eeq
we can say that the ensemble of triangulated manifolds exhibits
a simple {\it fractal} structure with Hausdorff dimension $d_H$.
It turns out that direct fits at fixed volume to eqn.~\ref{dist} are
unable to reveal a convincing power law regime because of the
presence of large finite volume corrections. In the light of this
we have used a finite size scaling ansatz
to extract the physics hidden in $n\left(r,V\right)$ \cite{scaling}.
Specifically, we assume that $n\left(r,V\right)$ can be written 
in the form
\beq
n\left(r,V\right)=V^{1-\frac{1}{d_H}}f\left(\frac{r}{V^{\frac{1}{d_H}}}\right)
\eeq

\begin{figure}[htb]
\vspace{9pt}
\epsfxsize=2.8 in
\epsfbox{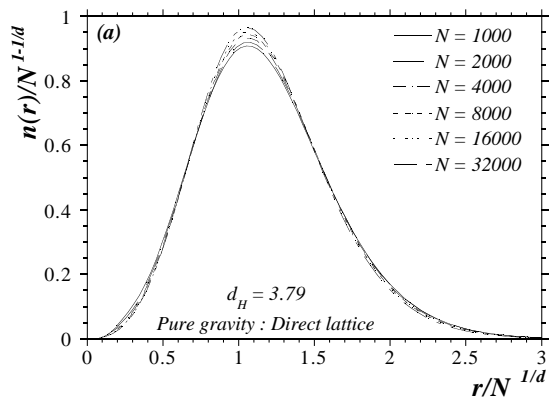}
\caption{Scaling plot for $n\left(r,V\right)$ for pure gravity}
\label{fig1}
\end{figure}

We show in fig.~\ref{fig1} a plot of the function
$f\left(r\frac{r}{V^{\frac{1}{d_h}}}\right)$ obtained from lattice sizes
varying from $V=500-32000$ in the case of pure gravity. Clearly
the data collapse almost perfectly onto a single scaling curve. The
optimal $d_H$ may be estimated by a suitable fitting procedure and
here corresponds to $d_H=3.8(1)$ - very consistent with a recent
analytic calculation which predicts $d_H=4$ 
\cite{ambwat}.
Notice that the numerical
calculations provide more than just a direct confirmation of this
fractal dimension - they show graphically the emergence of 
{\it nonperturbative} length scale in these quantum gravitational
systems which is power-like related to the total volume.
Similar numerical results have been reported by the Copenhagen group
\cite{ambscale}. 

\begin{figure}[htb]
\vspace{9pt}
\epsfxsize=2.8 in
\epsfbox{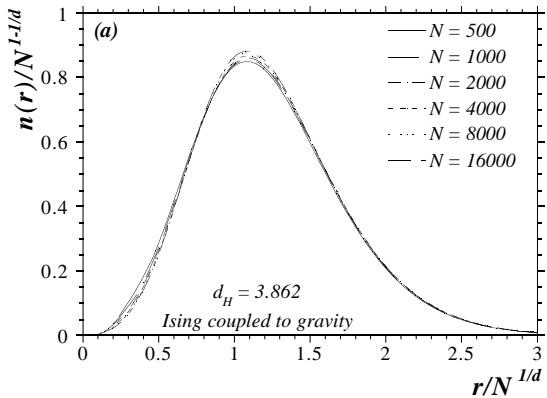}
\caption{Scaling plot for $n\left(r,V\right)$ for Ising+gravity}
\label{fig2}
\end{figure}

The numerical approach can be trivially extended to study $c>0$
theories coupled to quantum gravity. An example
of a scaling plot for the critical Ising system is
shown in fig.~\ref{fig2}. The conclusion from these studies
can be simply summarized; for $c<1$ no backreaction of the critical
matter field is visible in the measured Hausdorff dimension which
remains at four, for $c>1$ scaling is still observed with a
Hausdorff dimension which decreases monotonically towards two with
increasing $c$ - the value for branched polymers. These latter
results were obtained in \cite{ambscale} together with
corresponding results for another fractal dimension - the spectral
dimension.

Preliminary results in four dimensions have indicated that a
similar scaling behavior may also occur there and constitutes
further evidence for critical behavior in the model \cite{smitscal}.

\subsection{MCRG for 2d gravity}

Much of the numerical work in 2d gravity
has centered around the application of
finite size scaling to extract critical exponents see eg \cite{generic_num}.
In conventional statistical mechanical systems the 
validity of such techniques depends on the existence of
a renormalisation group (RG) which governs the flow
of couplings under changes in scale. It is tempting to
believe that a similar RG structure must underlie the DT models
and is responsible for the observed scaling.

In the continuum formulations of quantum gravity the very issue
of a renormalisation group is a difficult one to formulate since the theory
(in the absence of a cosmological constant) possesses no length
scale. In contrast the DT formulation contains an invariant cut-off
corresponding to the elementary triangle edge length. The latter
may be traded in for the number of triangles $N$ if the physical
volume is held fixed.

A successful RG transformation would be important both conceptually
and as a powerful new tool with which to compute critical
points and critical exponents for systems 
coupled to quantum gravity. We have undertaken a
systematic exploration of the
features of one such 
blocking transformation used in conjunction with Monte Carlo
simulation \cite{rg}.

In conventional lattice field theory a RG or block transformation acts so
as to reduce the number of degrees of freedom by replacing the
original (bare) theory by one defined on a coarser lattice. If this
new theory is to preserve the essential long distance physics there
will be an associated flow in the coupling constants of the model. Notice
that for the usual field theory models on regular (flat space) lattices
the blocking transformation trivially preserves the lattice
structure and the main problem is to devise a blocking
scheme for the matter fields.

Quantum gravity is quite different -- it is not possible to
exactly preserve the features of any given fine triangulation under
blocking. Indeed the lattice itself now carries dynamical degrees
of freedom. Thus the transformation we envisage will replace
a given triangulation $T\left(N\right)$ by some 
triangulation $T^\prime\left(N^\prime\right)$ corresponding to
a blocking factor $b=N/N^\prime <1$.
\beq
T^\prime\left(N^\prime\right)=R\left(T\left(N\right)\right)
\eeq
Clearly the choice of the transformation $R$ is of crucial
importance. Presumably it should  have the property of preserving certain
aspects of the long distance geometry. Recently various 
proposals have been put forward for accomplishing this in
the case of pure gravity. The first of these
approaches due to Renken \cite{renken} associates nodes of the
block lattice with a randomly selected subset of the fine lattice
points. The block triangulation is then chosen so as to
preserve {\it locally} the relative geodesic distances of these block
nodes.

Another approach advocated by Krzywicki et al. \cite{krz} constructs
a block triangulation by eliminating a class of extremal baby
universes associated with the original random lattice. For
a discussion of baby universes see later. Since the
baby universe distribution is intimately connected to the
fractal structure of the ensemble of triangulated manifolds it is
hoped that this transformation will (approximately) preserve the fractal
structure. This approach has recently been applied to four
dimensions resulting in a prediction for the $\beta$-function 
corresponding to the node coupling $\kappa_0$ in the vicinity
of the transition \cite{burda}. However it remains to be shown 
whether this method can yield the correct gravitationally
dressed exponents in two dimensions.

Here we have examined another blocking method based on
a {\it local} direct decimation of the initial lattice. The algorithm
is
extremely simple. First pick a point at random in the fine lattice. If
we attempt to remove this node then unless the
original node coordination was three we will be left with a polygonal
`hole'
in the triangulation. However by the addition of suitable links it
is possible to retriangulate the interior of this hole. We choose
one of the many possible retriangulations at random.

By iterating this procedure an arbitrary number of
times we can produce a block lattice of any volume.
In practice, this procedure is effected by randomly flipping the
links around a selected node until its coordination number
becomes equal to three. At this point any original curvature 
associated with the node has been smeared out over its neighbors. We then
remove the three-fold node. Thus this method can be seen to
closely preserve the local curvature.  

\begin{figure}[htb]
\vspace{9pt}
\epsfxsize=2.8 in
\epsfbox{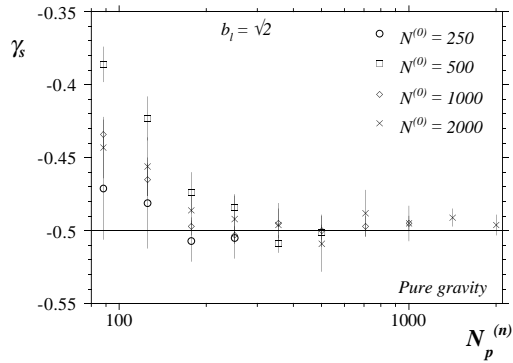}
\caption{$\gamma_{str}$ vs blocking level for pure gravity}
\label{fig3}
\end{figure}

\begin{figure}[htb]
\vspace{9pt}
\epsfxsize=2.8 in
\epsfbox{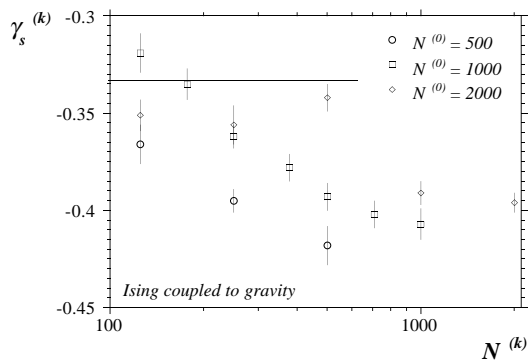}
\caption{$\gamma_{str}$ vs blocking level for Ising+gravity}
\label{fig4}
\end{figure}

In the case when a 
spin configuration decorates the original lattice we have
experimented with two methods for blocking the lattice. The first
proceeds as for pure gravity, the lattice blocking being
unaffected by the spin configuration. In the second method
we subject the link flipping to a standard metropolis test
using the critical Ising action. The former appears to give
more stable critical exponent estimates. The block spins
are simply arrived at by a variation on the majority rule:
a given block spin is determined by examining its neighbors on
the original lattice (excluding surviving block nodes) and
assigning a spin according to the sign of the sum of their spins.
I refer the reader to Thorleifsson's
talk at this meeting \cite{rg} and
reference \cite{thorleif} for further details and here merely show results for
the exponent $\gamma_{str}$ obtained after blocking, both for
pure gravity and gravity coupled to Ising spins figs.~\ref{fig3} and
\ref{fig4}.

\begin{figure}[htb]
\vspace{9pt}
\epsfxsize=2.8 in
\epsfbox{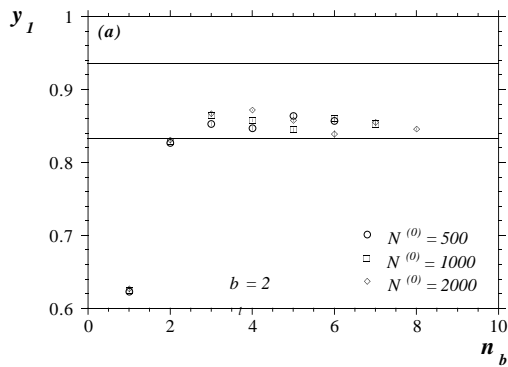}
\caption{$\frac{1}{\nu d_H}$ vs blocking level for Ising+gravity}
\label{fig5}
\end{figure}

\begin{figure}[htb]
\vspace{9pt}
\epsfxsize=2.8 in
\epsfbox{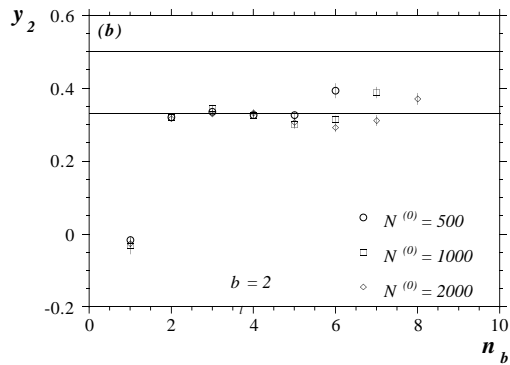}
\caption{$\frac{\delta}{\delta+1}$ vs blocking level for Ising+gravity}
\label{fig6}
\end{figure}

Clear evidence for fixed point behavior is seen with
the correct gravitational exponent. A more stringent test
can be made by looking at the dressing of the Ising exponents extracted
in the usual MCRG manner from an estimated $T$-matrix built
out of a truncated operator basis. Figure~\ref{fig5} shows estimates
of the Ising exponent $\frac{1}{\nu d_H}$ and fig.~\ref{fig6} the
exponent
$\frac{\delta}{\delta+1}$ for the critical
Ising plus gravity system as a function of blocking level. 

These
numbers correspond to an even spin operator basis
consisting of the simple energy bond operator, the
conjugate bond operator and the product of the four spins
around the boundary of two neighbour simplices. The odd
spin operator basis is formed from the simple single spin
operator and the product of spins around a triangle.

The theoretical prediction governed by the
KPZ formula is shown by the lower horizontal line while
the solid line above indicates the Onsager flat space limit. Very
encouragingly we see the numerical results are both stable under
blocking and close (in most cases statistically consistent with) the
KPZ predictions. This is highly non-trivial of the 
correctness of our blocking procedure. 

It is clearly important to check the efficacy of this blocking
scheme on other solved gravity coupled models and to extend it
to the case of scalar fields. One of the prime advantages this
method possesses over the usual finite size scaling is that
the critical coupling is an end result of the calculation - it is
not needed as input to determine critical exponents. Both $c>1$
models and the crumpling transition furnish examples of suitable
arenas for application of these ideas. 

\section{Recent Progress in Four Dimensions}

\subsection{Critical Properties}

All the groups that have studied this model report evidence for
two phases separated by a continuous phase transition. For small node
couplings $\kappa_0$ the model is in a crumpled phase with a typical
mean intrinsic extent growing only logarithmically with volume and
negative total curvature. For large $\kappa_0$ the model enters a
branched polymer phase characterized by a mean size varying as
the square root of the volume (Hausdorff dimension two) and
positive mean curvature. A phase transition separates these
regimes revealed by a node susceptibility whose peak value
scales in power-like fashion with the volume.

These calculations have been greatly strengthened in the past year
by a calculation of Ambj\o rn and Jurkiewicz \cite{amb4d}
in which a new
algorithm was used to extend the range of lattice sizes to much
larger values than previously ($64K$ simplices). The basis of their
global algorithm termed `baby universe surgery' resides 
in the identification of minimal neck baby universes. Once one of
these has been located it may be moved globally to another region
of the manifold by randomly selecting a simplex on the `mother'
universe and gluing the baby at that point by identifying
the minimal neck with the boundary of that simplex. The authors
show that this procedure is very efficient in reducing
critical slowing down associated with the geometry near the
phase transition and within the branched polymer phase. The
location of baby universes also allows them to measure the
baby universe distribution and hence the exponent $\gamma_{str}$. 

The computed value is consistent with $\gamma_{str}=\frac{1}{2}$
throughout the branched polymer phase. This provides yet
another clue that the dominant configurations are indeed true
branched polymers. An analysis of the geodesic path
distribution, defined similarly as in two dimensions, shows
very accurate scaling throughout this region and provides
another accurate estimate for the Hausdorff dimension $d_H=2$.

By tracking the deviation of the exponent $\gamma_{str}$ from 
one half it is also possible to read off a new estimate for the
pseudo-critical coupling $\kappa_0^c\left(V\right)$
at finite volume. It is found that
\beq
\kappa_0^c\left(V\right)-\kappa_0^c\left(\infty\right)\sim V^{-\delta},
\eeq
with $\delta=0.47(3)$. This is a significant improvement over
past estimates which have estimated this pseudo-critical point by
following the volume dependence of the peak in the node
susceptibility.

Finally, by fitting the geodesic path distribution to a suitable
ansatz in the crumpled phase of the model it is possible to
extract a `massgap' $\Delta m\left(\kappa_0\right)$ for the model.
As the phase transition is approached it is found that
\beq
\Delta m\sim \left(\kappa_0 -\kappa_0^c\right)^\epsilon,
\eeq
with critical exponent $\epsilon=0.50(1)$

Baby universes have also been used in another way to analyze the
critical properties of this 4d phase transition, In \cite{burda} Burda,
Kownacki and Krzywicki employed an extension of their blocking
method for two dimensional quantum gravity to four dimensions. The
essence of the method is to identify baby universes of a certain 
size and remove them from the triangulated manifold closing the
minimal neck with a simple simplex and discarding the simplices 
associated with the original `baby'. This reduction of the volume
can be traded for a change in length scale and hence an RG
transformation. By reasoning that this elimination of extremal
baby universes at least approximately preserves the fractal
structure of the ensemble of manifolds it it hoped that it
is a truly `apt' transformation. The authors are able to
compute the beta-function for the node coupling $\kappa_0$ close
to the critical point and indeed observe a IR unstable zero
very close to the peak in the node susceptibility. This provides
further confirmation of the transition and also indicates that
the fixed point is not, as has been recently speculated \cite{anton},
an IR stable fixed point associated with the conformal mode. 

\subsection{Two-Point Functions}

The last year has also witnessed substantial progress in the
understanding of two-point correlation functions in 4d gravity.
De Bakker and Smit argue that the definition of a `connected'
correlator is somewhat subtle in the situation where a sum
over geometries is being performed \cite{smit}. They present
numerical results which support the idea that it is necessary to
subtract a {\it distance dependent} one-point function to
obtain a simple behavior for the two-point function.

One way to understand this procedure is to consider the definition
of a connected function down geodesic paths
\beq
C_Q\left(r\right)=\left\langle\frac{1}{n\left(r\right)}\sum_{ij}
\delta Q_i\delta Q_j \delta\left(
d_{ij}-r\right)\right\rangle,
\label{corr}
\eeq	
where
\begin{eqnarray}
n\left(r\right)& = &\sum_{ij}\delta\left(d_{ij}-r\right)\\
\delta Q_i &=& Q_i-\overline{Q}.
\end{eqnarray}

The quantity $Q_i$ can be any local operator associated with
simplex $i$ and $d_{ij}$ is the (dual lattice) geodesic distance
between simplices $i$ and $j$. Smit et al. consider the case where
$Q$ is the curvature density. In flat space the
expectation values $\overline{Q}$ are just constants; however in order
that eqn.~\ref{corr} factorizes at large distance to a form
$\left\langle Q_i Q_j\right\rangle -{\overline{Q}}^2$ it is 
necessary to allow the expectation value $\overline{Q}$ to depend on distance
down the geodesic path $r$. Specifically,
\beq
\overline{Q}=\overline{Q}\left(r\right)=\left\langle\frac{1}{n\left(r\right)}
\sum_{ij}Q_i\delta\left(d_{ij}-r\right)\right\rangle
\eeq
If the operator is completely uncorrelated with the
geometry (the situation in flat space for example) this expression
reduces to the usual one. However, in general, when we sum
over triangulations it is perhaps natural to assume that
the detailed distribution of matter field configurations will
depend on geometry and the average of some operator on
geodesic circles may not be the same as its bulk
expectation value.
This appears to be the case and data is shown illustrating that
when such a subtraction is made the correlation function
simplifies to yield a simple power law behavior.
Indeed close to the transition it appears that
\beq
C_R\left(r\right)\sim \frac{1}{r^4}
\eeq
which is consistent with the exchange of two gravitons.

\subsection{Open Issues}

Over the previous two years there has been a considerable amount of
activity devoted to the question of an exponential bound in 
DT quantum gravity. The issue is this; does the number of
triangulations of, for example, the d-sphere, grow 
no faster than exponentially with the volume. In two dimensions
this is known to be true and is the basis for taking the
continuum limit - the bare cosmological constant is tuned to
compensate for this exponential growth leading to a power
divergence of the mean volume and susceptibility.

\begin{figure}[htb]
\vspace{9pt}
\epsfxsize=2.8 in
\epsfbox{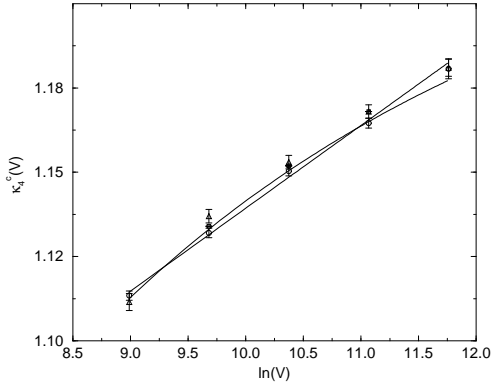}
\caption{$\kappa_4^c\left(V\right)$ vs $\log{V}$ at $\kappa_0=0$}
\label{fig7}
\end{figure}

In higher dimensions there are no rigorous proofs of such
a bound and one must turn to numerical simulation to
attempt to answer this important question. 
In three dimensions the results of these studies have
been pretty convincing \cite{me3,amb3} - at large volumes
the partition function indeed saturates at exponential
growth.
However in four dimensions the results have been more
controversial. In \cite{prl} evidence supporting an
unbounded scenario was presented. However the results
of simulations at larger volumes were shown to be
consistent with such a bound \cite{enz_bound,amb_bound}.
Figure~\ref{fig7} hows a plot of the effective
critical cosmological constant $\kappa_4^c\left(V\right)$ at
node coupling $\kappa_0=0$ with results from the three
groups working on this issue. 

A bound would be indicated by a effective
coupling that becomes independent of volume at
large volume. This is not seen at the
lattice volumes accessible to experiment ($V\le 128000$ simplices
here). So the question of the bound translates into
whether suitable empirical fits to the data favor
a coupling which becomes constant for infinite
volume or merely keeps increasing.

Two scenarios are usually used for the fits; a
logarithmic divergence of $\kappa_4^c\left(V\right)$ with
volume or a weak power law convergence. The issue is simply
which fit describes the data best and how robust is that fit - 
how confident are we that the fit is stable over a wide
range in volume. 

The first point to note is that the data from different
groups are all statistically consistent with each other. The straight
line and curve represent the best fits to our data assuming
the log and power scenario respectively. 
The quality of the
fits as revealed in the chi square per degree of freedom are
comparable at order unity and our conclusion must be
that it is {\it impossible} to distinguish confidently
between the two
possibilities from this data alone.  

Faced with this problem we have turned to the analysis of
other quantities in the hope of resolving the issue. One such
quantity is the distribution of baby universes. A baby universe
is defined as a portion of the triangulation which is separated from
the bulk by a so-called `minimal neck'. In $d$ dimensions the
latter is defined to be $d+1$ $(d-1)$-simplices which form the
boundary of a $d$-simplex {\it not} already present in the
triangulation. The fraction of such baby universes with volume
$B$ is a sensitive test of the subleading behavior of the
microcanonical partition function and hence yields information
on the exponential bound.

We have conducted a careful study of this distribution measured
on lattices with volumes $V=500-8000$ and fitted the data to
one of the two following forms corresponding to logarithmic
divergence and power law convergence.

\begin{eqnarray}
\log{P\left(B\right)}&=&a+\beta \left(B+\delta\right)
\log{\left(B+\delta\right)}+\nonumber\\
                     & &\beta 
\left(V-B+\delta\right)\log{\left(V-B+\delta\right)}
\end{eqnarray}

\begin{eqnarray}
\log{P\left(B\right)}&=&a^\prime+\beta^\prime \left(B+\delta\right)^
{1-\gamma}+\nonumber\\
& &\beta^\prime \left(V-B+\delta\right)^{1-\gamma}.
\end{eqnarray}

The constant $\delta$ is inserted as a phenomenological parameter to
reflect sub-leading finite size corrections and $a$ and $a^\prime$
reflect an ambiguity in overall normalization. In practice we
have removed the largest contribution to the latter by
dividing the measured number of baby universes by the volume $V$

\begin{figure}[htb]
\vspace{9pt}
\epsfxsize=2.8 in
\epsfbox{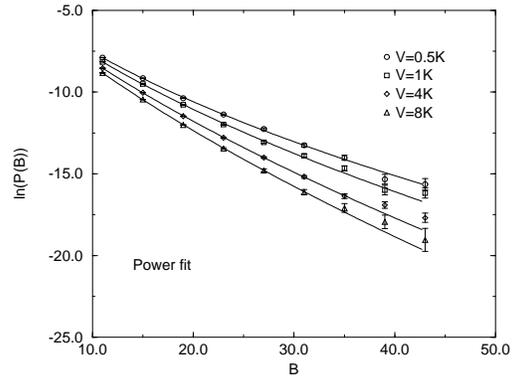}
\caption{$\log{P(B)}$ vs $B$ with power fit}
\label{fig8}
\end{figure}

Figure 8 shows the data fitted according to the power
scenario Eq.\ (7). The best fit in this case yields
$a^\prime=-0.2(15)$, $\beta^\prime=-1.38(5)$ and $\delta=3(2)$ with
$\chi^2=6.3/6$ assuming $\gamma=0.25$ as before. The log
gives a marginally worse fit with $\beta=0.056(1)$ but still
is consistent with the data.
At face value it remains hard to differentiate between the
two situations. However, notice that the extracted value
of $\beta=0.056(1)$ from the log fit is more than twice its 
estimated value from the fits for the effective critical
coupling $\beta=0.025(1)$. 
In contrast the estimate for $\beta^\prime=-1.38(5)$ from
the power fit is quite close to its value estimated from the
critical coupling (figure 7)
$\beta^\prime=-1.23(4)$. The relative proximity of the
two estimates is particularly impressive considering that
one is derived from the behavior of baby universes with
size less than $8000$ simplices while the other is extracted
from the critical coupling at volumes much greater than $8000$.
Furthermore, it is clear that the power fit would still hold good
if we set $\delta=a^{\prime} =0$ so that such a fit (with a truly
minimal number of parameters) would do much better than the logarithm.

In conclusion, the numerical results which have been
obtained by three groups, although not
definitive, are very consistent with the
existence of an exponential bound in the dynamical 
triangulation model of $4d$ quantum gravity. The evidence for this
comes both from fits to the volume dependence of the critical coupling, 
and an analysis
of the baby universe distribution in the crumpled phase. 

Although individually these quantities are not very
conclusive, it is remarkable how consistent results are
obtained if we assume a weak power convergence.
Clearly, it is important
to strengthen these conclusions both by simulating intermediate
lattice volumes and perhaps via a high statistics simulation at
say volume $V=16000$ directed at probing further into the tail of
the baby universe distribution.

Finally I should like to mention the work of the Japanese group
\cite{japan} who have observed that `singular' vertices are
present in four dimensional DT. These are vertices which
are common to a fixed fraction of the total number of simplices.
There appear to be two such vertices in 4d whose distribution of
simplex coordination is well separated from the `bulk'
distribution corresponding to all other vertices in the
triangulation. I refer the interested reader to the 
parallel session given at this meeting for further details and
merely make two remarks here.
Firstly, it appears that the dynamics of these singular
vertices plays an important role in the longest autocorrelation
times observed for these systems \cite{ergod}. Furthermore,
a crude suppression of these objects appears to remove the
phase transition. 
Secondly, no such vertices exist in three dimensions which is
probably the first qualitative indication that the physics of
three-dimensional and four-dimensional DTs can be quite
different. 

I would like to acknowledge the support of Syracuse University
during the course of this work and useful conversations
with Gudmar Thorleifsson, Ray Renken and John Kogut.

\end{document}